\documentclass[aps,preprint,superscriptaddress,nofootinbib]{revtex4}%
\usepackage{amsfonts}
\usepackage{amsmath}
\usepackage{amssymb}
\usepackage{graphicx}
\newcommand{\be}{\begin{equation}}
\newcommand{\ee}{\end{equation}}
\newcommand{\bea}{\begin{eqnarray}}
\newcommand{\eea}{\end{eqnarray}}

\begin{document}

\title[HGT and (2+1)-dimensional gravity]
    {\textbf{HIGHER GAUGE THEORY AND GRAVITY IN (2+1) DIMENSIONS}}

\author{\textbf{R.\ B.\ Mann}}

    \email{mann@avatar.uwaterloo.ca}

    \affiliation{Perimeter Institute for Theoretical Physics, Waterloo,
                 Ontario, Canada, N2J 2G9}

    \affiliation{Department of Physics, University of Waterloo, Waterloo,
                 Ontario, Canada N2L 3G1}

\author{\textbf{Eugeniu M.\ Popescu}}

    \email{empopesc@uwaterloo.ca}

    \affiliation{Department of Physics, University of Waterloo, Waterloo,
                 Ontario, Canada N2L 3G1}


\begin{abstract}
\noindent
        Non-abelian higher gauge theory has recently emerged as a generalization
        of standard gauge theory to higher dimensional (2-dimensional in the
        present context) connection forms, and as such, it has been successfully
        applied to the non-abelian generalizations of the Yang-Mills theory and
        2-form electrodynamics. (2+1)-dimensional gravity, on the other hand, has
        been a fertile testing ground for many concepts related to classical and
        quantum gravity, and it is therefore only natural to investigate whether
        we can find an application of higher gauge theory in this latter context.
        In the present paper we investigate the possibility of applying the
        formalism of higher gauge theory to gravity in (2+1) dimensions, and we
        show that  a nontrivial model of (2+1)-dimensional gravity coupled to
        scalar and tensorial matter fields - the $\Sigma\Phi EA$ model - can be
        formulated both as a standard gauge theory and as a higher gauge theory.
        Since the model has a very rich structure - it admits as solutions
        black-hole BTZ-like geometries, particle-like geometries as well as
        Robertson-Friedman-Walker cosmological-like expanding geometries - this opens
        a wide perspective for higher gauge theory to be tested and understood in
        a relevant gravitational context. Additionally, it offers the possibility of
        studying gravity in (2+1) dimensions coupled to matter in an entirely new
        framework.

\end{abstract}

\date[Date: ]{July $31$, 2006}

\startpage{1}
\endpage{25}

\maketitle


\newtheorem{definition}{Definition}


\section{Introduction}

\indent
    Higher gauge theory\footnote{Although we will be using in all of the following the
    term "higher gauge theory" to conform to the terminology already in use in the
    literature, we are in fact referring to "2-gauge" theory, as it will become obvious
    shortly.} (HGT), in its non-abelian version, has emerged in the last few years as
    the natural generalization - from both the mathematical and the physical perspective
    -  of the more traditional standard gauge theory (SGT) to consistently include
    non-abelian higher 2-form connections. Alternatively, HGT can be viewed, in the
    integral formulation, as the non-abelian generalization to "surfaces" of the SGT,
    which involves only curves.\\
\indent
    It is not our intention to develop or present general arguments in support of
    the above statement. Excellent such arguments are already extant in the literature
    \cite{Baez1}, \cite {PfGir}. Instead, we will focus on illustrating how the concept
    of a HGT can be applied in the context relevant for this paper - namely in
    (2+1)-dimensional gravity - and on exploring its implications.\\
\indent
    For this purpose, we will begin by considering pure gravity in (2+1)-dimensions
    (with vanishing cosmological constant) as an $SO(2,1)$ gauge theory. The action
    of the theory is given by the expression:
        \be
            S[E,A]=\int_{M}Tr\big(E\wedge R[A]\big)\label{eqn1}
        \ee
    where $M$ is the spacetime manifold, $E$ and $A$ are $so(2,1)$ Lie algebra valued
    1-forms associated with the triad\footnote{The $E$ fields can be identified with the
    spacetime triads only if they are invertible, but in the following, for simplicity
    reasons, we will assume that the $E$ fields are in fact invertible and we will refer
    to them as the triads.} fields and the Lorentz connection respectively,
    $R[A]=dA+A\wedge A$ is the curvature 2-form of the Lorentzian connection $A$, and
    $Tr(...)$ is the non-degenerate invariant bilinear form that can be defined on the
    $so(2,1)$ Lie algebra. It is well known that the theory has only topological degrees
    of freedom and as such it is exactly solvable \cite{witten1},\cite{horowitz1}.\\
\indent
    The first order variation of the action (\ref{eqn1}) yields the equations of motion
    $R[A]=DE=0$, where "D" stands for the covariant derivative of the triad field. These
    equations of motion state in fact that the spin-connection is locally flat ($R[A]=0$),
    and that the spacetime has vanishing torsion ($DE=0$). Assuming now that the spacetime
    has a topology $M=R\times S$ with $S$ a spacelike surface, the equations of motion
    split canonically into evolution equations for the fields (the components of the
    equations that involve the time derivatives of the fields) and into constraints (the
    components of the equations involving only the spatial derivatives of the fields), the
    latter becoming the generators of the gauge symmetries of  the theory. The constraints
    resulting from the flatness condition become the generators of the Poincar\'{e}
    translational symmetries, while the constraints resulting from the vanishing of the
    torsion become the generators of Lorentz symmetry transformations. The Poisson algebra
    of these constraints is the Poincar\'{e} algebra, and hence they generate the
    Poincar\'{e} ($ISO(2,1)$) group as the gauge group for (2+1)-dimensional gravity.\\
\indent
    Under these circumstances, one can immediately construct a very important quantity
    for the theory, namely the holonomy of the spin-connection $A$ along curves
    embedded in the spacetime manifold. The importance of the holonomy resides in the
    fact that on one hand, for closed curves, it provides - upon tracing - observables
    for the theory due to its invariance under the action of the Poincar\'{e} gauge
    group, and on the other hand - and this is the aspect of relevance for the present
    discussion - it provides a natural way for labeling curves and loops
    in the spacetime manifold. It is quite straightforward to understand this latter
    issue of labeling. As a path-ordered product of exponentials of $so(2,1)$ Lie
    algebra elements, the holonomy of the spin connection is itself an $SO(2,1)$
    group element, and as such it can be naturally associated with its underlying
    curve. In this way, every curve embedded in the spacetime manifold can be labeled
    by the $SO(2,1)$ group element corresponding to the holonomy of the spin connection
    along the curve, as one would have expected to be the case for a SGT. So the picture
    of (2+1)-dimensional gravity as an SGT is quite clear. \\
\indent
    The next issue that one has to consider is the generalization of (2+1)-dimensional
    gravity to the framework of HGT. According to the fundamentals of HGT, in the
    differential picture such a generalization should involve - in addition to
    the Lorentzian connection $A$ - a 2-form field that satisfies the requirements
    of a 2-connection, i.e. a 2-connection whose 3-curvature must satisfy certain
    mathematical conditions. Since such a 2-connection can naturally be integrated
    over a 2-dimensional manifold (over a surface), in the integral picture the
    holonomy of the 2-connection would offer a natural way for the surfaces enclosed by
    curves to also be labeled in a consistent way by group elements.\\
\indent
    Our approach to the generalization of (2+1)-dimensional gravity to the HGT
    framework is based on the main idea that the HGT formalism should reduce
    to the classical SGT formalism in the limit of a vanishing connection
    2-form. Such a generalization, while not universal - one could devise other
    has ways to generalize gravitation in (2+1) dimensions in the present
    context - has several practical advantages, as will be discussed in more detail
    below. Also, from the principial viewpoint, such a generalization is justified
    by the fact that it is the most natural and consistent way of extending a known
    theory to a more general framework.\\
\indent
    With the above considerations, we can immediately make a few very important
    observations that will help us with the construction of the generalized model.
        \begin{enumerate}
            \item Under the assumption that the HGT generalization of (2+1) dimensional
            gravity reduces to the classical SGT formalism, the limit
            of vanishing 2-connection implies that the Lagrangian density of the action
            of the generalized model should necessarily contain the term
            $Tr(E\wedge R[A])$. Under these circumstances the 2-connection should enter
            the generalized Lagrangian in a separate term. It is now rather obvious why
            we have chosen this approach for the generalization of (2+1)-dimensional
            gravity to the HGT formalism. The term $Tr(E\wedge R[A])$ present in
            the in the generalized lagrangian will yield upon first order variation of
            the fields equations of motion involving the 2-curvature of the lorentzian
            1-connection and the torsion of the triad, and as such, it will allow one
            to determine the geometry of the spacetime - in principle at least - directly
            from the equations of motion of the generalized model. Of course, these
            equations of motion will be more complicated than their pure gravity
            counterparts since they will contain additional terms involving the
            2-connection whose explicit form will be dependent on how the 2-connection
            is actually coupled to pure gravity. Nevertheless, they will offer a clear
            picture of how the spacetime geometry will affected by the presence of the
            2-connection of the HGT formalism.
            \item By introducing the 2-form connection in the generalized action, we
            have in fact introduced a new canonical variable, and if we want the classical
            formulation to be consistent, it is absolutely necessary to introduce an
            additional field/variable, which should be canonically conjugate
            to the 2-connection. Furthermore, since canonical conjugacy implies the
            presence of an exterior derivative operator acting on one of these variables,
             a simple counting argument shows that the variable conjugate
            to the 2-connection can only be a 0-form.
            \item We have established that the additional term that has to be added to
            the Lagrangian density of pure gravity in order to generalize the latter
            to the framework of HGT has to be a functional of a 2-form and a 0-form, and it
            must include at the very least an exterior derivative operator acting on one
            of these two forms. However, since we want the 2-form to be a 2-connection,
            and not some arbitrary field, the structure of this additional term must
            also be such that upon a first order variation of the fields it should yield
            an equation of motion that involves the 3-curvature of the 2-form. Under
            these circumstances, an obvious choice for the term to be added to the
            Lagrangian density of pure gravity - and by analogy with the latter -
            is a functional linear in the 3-curvature of the 2-connection and also
            linear in the conjugate 0-form field.
        \end{enumerate}

\indent
    With these observations, we now have a quite clear picture of the theory that
    generalizes pure (2+1)-dimensional gravity to the formalism of HGT. If $\Phi$
    are 0-form Lie algebra valued fields and $\Sigma$ is the Lie algebra valued
    2-connection with the 3-curvature given by the expression:
        \be
            G[\Sigma]=d\Sigma +A\rhd\Sigma\label{eqn2}
        \ee
    where $"\rhd"$ is an action of (Lie) algebras that will be discussed in more
    detail in the next section, then the action for (2+1)-dimensional gravity in
    the HGT formalism has the general form:
        \be
            S_{HGT}=\int_{M}\big\{Tr_{1}\big(E\wedge R[A]\big)+
                    Tr_{2}\big(\Phi\wedge G[\Sigma]\big)\big\}\label{eqn3}
        \ee
    Of course, there are several other issues that must be made explicit, like the
    underlying 2-algebra structure of the fields and the two traces in (\ref{eqn3}),
    but nevertheless, under the assumptions of our approach, (\ref{eqn3}) represents
    the most natural generalization of pure gravity in (2+1)-dimension to the
    formalism of HGT.\\
\indent
    At this time, there is one more and extremely important observation that we
    need to make. The expression (\ref{eqn3}) of the generalized action is
    reminiscent of a model that has already been studied in the literature, namely
    the so-called $\Sigma\Phi EA$ model \cite{mannpop} whose action is given by the
    expression\footnote{The original action of the $\Sigma\Phi EA$ model differs
    from the action in (\ref{eqn4}) by a surface term, but this is irrelevant for
    the content of the present paper}:
        \be
            S_{\Sigma\Phi EA}=\int_{M}Tr\big\{E\wedge R[A]+
                        \Phi\wedge D\Sigma\big\}\label{eqn4}
        \ee
    where in (\ref{eqn4}) $\Phi$, $\Sigma$ are 0-form fields and 2-form fields
    respectively, the trace is the non-degenerate invariant bilinear form
    defined on the Lie algebra of $SO(2,1)$ and $"D"$ is the classical exterior
    covariant derivative.\\
\indent
    The resemblance between the two models is not merely a matter of
    appearance. As we will show, the $\Sigma\Phi EA$ model can be
    reformulated as a higher gauge theory over the Poincar\'{e} 2-algebra, with
    an action having exactly the form (\ref{eqn4}). As such, and since it has
    non-trivial solutions like the BTZ black-hole geometry and the point-particle
    geometry, the $\Sigma\Phi EA$ constitutes the sought generalization of pure
    gravity in (2+1) dimensions to the HGT formalism, and to the best knowledge
    of the authors, the first non-trivial application of the HGT formalism to
    gravity.\\
\indent
    The remainder of the paper is organized as follows. In Section II and
    Section III we review the basic notions of the HGT formalism and of the
    $\Sigma\Phi EA$ model respectively, as they pertain to the purpose of this
    paper. In Section IV we show explicitly how the $\Sigma\Phi EA$ model can
    be reformulated to become the generalization of pure gravity in (2+1)
    dimensions to the HGT formalism, and in Section V we provide a
    discussion of the results and some concluding remarks regarding the
    future applications of the HGT formalism in gravity.

\section{2-groups, 2-algebras and higher gauge theory}

\indent
    In this section, we will present the fundamentals of the HGT formalism, limiting
    ourselves only to those concepts that are relevant for the contents of this paper
    for reasons of simplicity. Under these circumstances, the presentation of these
    concepts is by no means self-contained, and for more details, the interested
    reader is referred to \cite{Baez1}, \cite{PfGir}, \cite{pfeiffer1} and the
    references therein.

\subsection{Lie 2-groups and Lie 2-algebras}

\indent
    The relevant mathematical structure underlying the HGT formalism in the integral
    formulation is the "strict" (Lie) 2-group, which can be viewed as the generalization
    of the (Lie) group structure underlying the classical SGT. The beauty of the HGT
    formalism is that due to its categorical nature, there are several equivalent ways
    in which a "strict" 2-group can be defined \cite{lauda1}, \cite{for-bark}, and
    consequently one can choose the definition that best fits one's purpose. In our
    particular case, the  definition best fitted for a "strict" 2-group is that of a
    crossed module, and we will give this definition below without insisting on its
    explicit relation with the (original) categorical definition beyond the statement
    that the two definitions are equivalent.
        \begin{definition}
            A Lie 2-group as a crossed module is a quadruple $(G,H,t,\diamond)$
            consisting of:
                \begin{itemize}
                    \item[\textbf{a.}]
                        Two Lie groups $G$, $H$.
                    \item[\textbf{b.}]
                        A smooth group homomorphism $t:H\rightarrow G$, i.e. a map $t$
                        satisfying the relations:
                            \bea
                                t(h_{1}\cdot h_{2})&=&t(h_{1})\cdot t(h_{2})\nonumber\\
                                t(1_{H})&=&1_{G}\label{eqn5}
                            \eea
                        for any $h_{1}, h_{2}\in H$.
                    \item[\textbf{c.}]
                        A smooth action $"\diamond"$ of $G$ on $H$ by automorphisms
                        (a smooth homomorphic map $G\rightarrow Aut[H])$, i.e. a group
                        action satisfying the relations:
                            \bea
                                (g_{1}\cdot g_{2})\diamond h&=&
                                    g_{1}\diamond( g_{2}\diamond h)\nonumber\\
                                1_{G}\diamond h&=&h\label{eqn6}\\
                                g\diamond (h_{1}\cdot h_{2})&=&
                                    (g\diamond h_{1})\cdot(g\diamond h_{2})\nonumber\\
                                g\diamond 1_{H}&=&1_{H}\nonumber
                            \eea
                        for any $g, g_{1}, g_{2}\in G$ and any $h, h_{1}, h_{2}\in
                        H$.
                \end{itemize}
            The map $t$ and the operation $"\diamond"$ are required to satisfy the
            following two sets of compatibility conditions:
                \begin{itemize}
                    \item[\textbf{i)}]
                        For any $g\in G$ and for any $h, h'\in H$
                        we have:
                            \bea
                                t(g\diamond h)&=&g\cdot t(h)\cdot g^{-1}\nonumber\\
                                t(h)\diamond h'&=&h\cdot h'\cdot h^{-1}\label{eqn7}
                            \eea
                    \item[\textbf{ii)}]
                        For any $g\in G$ and for any $h\in H$ there exists $g'\in G$
                        such that:
                            \be
                                g'=t(h)\cdot g\label{eqn8}
                            \ee
                \end{itemize}
        \end{definition}

\indent
    The compatibility condition (\ref{eqn8}), which is related to the generalized
    notion of flatness and therefore plays a fundamental role in the HGT formalism,
    is a direct consequence of the construction of the "strict" Lie 2-group as a
    crossed module \cite{Baez1}. However, there is a simpler and more intuitive
    way to understand it: since according to the definition, $t(h)\in G$, then for
    any $g\in G$, the product $t(h)\cdot g$ is an element of $G$, or in other words,
    there is an element $g'\in G$ such that $g'=t(h)\cdot g$, which is exactly what
    the compatibility condition $(ii)$ above states. The conditions in
    (\ref{eqn7}) have the role of ensuring the compatibility\footnote{It should be
    noted that the two conditions in (\ref{eqn7}) are not independent. For example,
    by applying the homomorphism $t$ to the second equation in (\ref{eqn7}) and by
    making use of (\ref{eqn5}) and (\ref{eqn6}), it is straightforward to recover
    the first condition in (\ref{eqn7}). However, the presentation of the compatibility
    relation between the map $t$ and the operation $"\diamond"$ by two separate
    conditions is clearer and more convenient from the practical viewpoint.} between
    the homomorphism $t$ and the operation $"\diamond"$, or in other words, of
    ensuring the consistency of the labeling and composition of surfaces.\\
\indent
    Analogous to the case of standard Lie groups, one can associate with any "strict"
    Lie 2-group a "differential" object, called a "strict" Lie 2-algebra, which
    can be thought of in the present context as the generalization of the notion of
    Lie algebra to the formalism of higher gauge theory. Also, similar to the case
    of the Lie 2-groups, there are several equivalent ways in which a Lie 2-algebra
    can be defined. For our purposes the most appropriate definition is that involving
    the concept of differential crossed module (which in turn can be thought of as
    the "Lie algebra" of a Lie crossed module). Once again, we will only give the
    definition of the Lie 2-algebra as a differential crossed module without insisting
    on the relation between this definition and the original categorical definition
    beyond the statement that the two definitions are equivalent.
        \begin{definition}
            A Lie 2-algebra as a differential crossed module is a quadruple
            $(\textbf{g}, \textbf{h}, \tau, \rhd)$ consisting of:
                \begin{itemize}
                    \item[\textbf{a.}]
                        Two Lie algebras $\textbf{g}$ and $\textbf{h}$.
                    \item[\textbf{b.}]
                        A homomorphism of Lie algebras
                        $\tau:\textbf{h}\rightarrow \textbf{g}$, i.e. a map satisfying
                        the relation:
                            \be
                                \tau([Y_{1},Y_{2}])=[\tau(Y_{1}),\tau(Y_{2})]\label{eqn9}
                            \ee
                        for any $Y_{1},Y_{2}\in \textbf{h}$.
                    \item[\textbf{c.}]
                        An action $"\rhd"$ of $\textbf{g}$ on $\textbf{h}$ by derivations,
                        i.e. a bilinear operation satisfying the relations:
                            \bea
                                [X_{1},X_{2}]\rhd Y&=&X_{1}\rhd (X_{2}\rhd Y)-
                                                    X_{2}\rhd (X_{1}\rhd Y)\nonumber\\
                                X\rhd [Y_{1},Y_{2}]&=&[X\rhd Y_{1},Y_{2}]+
                                                    [Y_{1},X\rhd Y_{2}]\label{eqn10}
                            \eea
                        for any $X, X_{1}, X_{2}\in \textbf{g}$ and for any
                        $Y, Y_{1}, Y_{2}\in \textbf{h}$.
                \end{itemize}
            The map $\tau$ and the operation $"\rhd"$ must satisfy the following two
            sets of compatibility conditions:
                \begin{itemize}
                    \item[\textbf{i)}]
                        For any $X \in \textbf{g}$ and for any $Y, Y' \in \textbf{h}$
                        we must have:
                            \bea
                                \tau(X\rhd Y)&=&[X,\tau(Y)]\nonumber\\
                                \tau(Y)\rhd Y'&=&[Y,Y']\label{eqn11}
                            \eea
                    \item[\textbf{ii)}]
                        On a hypercubic lattice, for any $X_{1} \in \textbf{g}$ and for
                        any $Y \in \textbf{h}$ there exists $X_{1} \in \textbf{g}$ such
                        that:
                            \be
                                \tau(Y)=X_{2}-X_{1}\label{eqn12}
                            \ee
                \end{itemize}
        \end{definition}

\indent
    Similar to the case of 2-groups, the compatibility condition (\ref{eqn12}) arises
    from the construction of a Lie 2-algebra from a differential crossed module. It
    is straightforward to see that is the lattice counterpart of (\ref{eqn8}), and as
    such, it has a similar intuitive explanation and it plays an equally important
    role - in the differential picture - in the concept of generalized flatness.\\
\indent
    The relation between Lie 2-groups and Lie 2-algebras is similar to the relation
    between standard Lie groups and Lie algebras. Once again, we will not expound upon
    the details of this relation beyond stating that every Lie 2-group has a Lie
    2-algebra \cite{Baez1}, and that the maps and operations that enter the above
    definitions are formally related through the expressions $\tau=d(t)$ and
    $"\rhd"=d("\diamond")$ \cite{Baez1}, \cite{PfGir}. Under these circumstances,
    the Lie 2-algebras emerge intuitively as the "differentials" of Lie 2-groups in
    the neighborhood of the 2-group identity, which is pretty much what one would
    have expected.

\subsection{Higher gauge theory}

\indent
    Once we have introduced the necessary mathematical structure, we can now proceed
    with the setting of the HGT formalism. Since in all of the following considerations
    we will be more interested in the "differential" picture rather than in the
    "integral" one, we will mainly limit ourselves to use the latter for pictorial
    purposes only.\\
\indent
    As mentioned in the introduction, the general purpose of higher gauge theory is
    to allow for the possibility of labeling both curves and surfaces in a (spacetime)
    manifold by group elements, and hence develop - intuitively speaking - a gauge
    theory of curves and surfaces. In the integral picture, this is done as follows.
    Given a 2-group $(G,H,t,\diamond)$, we label the curves in the manifold by elements
    $g\in G$ and the surfaces of the manifold that are bound by the curves with elements
    $h\in H$. The composition of curve labels is done as in classical SGT, and in
    addition, one can develop in a consistent manner composition rules for surface labels.
    Using these composition laws for curve and surface labels, one can then generalize
    the classical SGT concepts of gauge transformations, gauge invariance, etc. to
    obtain a consistent gauge theory - a 2-gauge theory - whose description and symmetries
    are determined by the underlying 2-group structure \cite{Baez1}, \cite{PfGir}.\\
\indent
    Alternatively, in the differential description, the underlying structure of HGT
    is a Lie 2-algebra $(\textbf{g}, \textbf{h}, \tau, \rhd)$, and on this Lie
    2-algebra we define a $\textbf{g}$-Lie algebra valued connection 1-form $A$ and a
    $\textbf{h}$-Lie algebra valued connection 2-form\footnote{The notation
    that we use for the 2-connection is different from the traditional notation used
    in \cite{Baez1}, \cite{PfGir}. Instead of $B$ for the 2-connection, we denote the
    2-connection $\Sigma$ in order to make clear the relationship between the HGT
    formalism and the $\Sigma\Phi EA$ model} $\Sigma$ (the 2-connection mentioned in the
    introduction). Of course, since the connection $A$ is a 1-form, it can naturally
    be integrated along curves, and as such it can be used to label the curves in the
    (spacetime) manifold as in the case of the classical SGT. Similarly, since the
    connection $\Sigma$ is a 2-form, it can naturally be integrated on surfaces,
    providing the labels for the surfaces of the (spacetime) manifold.\\
\indent
    Let  $\{\gamma^{a}\}$ be the generators of the Lie algebra $\textbf{g}$, and in
    all of the following, lower case latin indices form the beginning of the alphabet
    will denote $\textbf{g}$-algebra indices. We define the covariant derivative of a
    $\textbf{g}$-valued form $V=V_{a}\gamma^{a}$ as:
        \be
            D_{A}V=dV+[A,V]=
                dV+A_{a}\wedge V_{b}[\gamma^{a},\gamma^{b}]\label{eqn13}
        \ee
    and with this definition, the SGT curvature of the 1-connection $A$ is given by
    the expression:
        \be
            R[A]=dA+\frac{1}{2}[A,A]=
                da+\frac{1}{2}A_{a}\wedge A_{b}[\gamma^{a},\gamma^{b}]\label{eqn14}
        \ee
\indent
    The generalized HGT curvature of the 1-connection $A$ is given by the expression:
        \be
            F[A]=R[A]+\tau(\Sigma)\label{eqn15}
        \ee
    and in the HGT formalism, the generalized curvature $F[A]$ must always be vanishing,
    i.e. one must always have $F[A]=0$. This vanishing of the generalized curvature
    should not be confused with the concept of flatness of the connection 1-from $A$ in
    SGT. They are different concepts. As can be seen from (\ref{eqn15}) the
    vanishing of the generalized curvature implies in the general case a non-flat
    1-connection with the SGT curvature given by the expression:
        \be
            R[A]=-\tau(\Sigma)\label{eqn16}
        \ee
\indent
    Similarly, if we let $\chi^{m}$ be the generators of the Lie algebra $\textbf{h}$,
    and we consider latin lower case letters from the end of the alphabet to be
    $\textbf{h}$-algebra indices, we can define the covariant derivative of
    an
    $\textbf{h}$-valued form $W=W_{m}\chi^{m}$ with respect to the $\textbf{g}$-valued
    1-connection $A$ through the expression:
        \be
            D_{A}W=dW+A\rhd W=
                dW+A_{a}\wedge W_{m}(\gamma^{a}\rhd \chi^{m})\label{eqn17}
        \ee
    Furthermore, in the HGT formalism we can also define a curvature 3-form (the
    3-curvature) of the 2-connection $\Sigma$ through the relation:
        \be
            G[\Sigma]=D_{A}\Sigma=d\Sigma+A\rhd \Sigma\label{eqn18}
        \ee
    and using this definition, we can introduce the concept of 2-flatness the
    3-connection to vanish.\\
\indent
    The infinitesimal gauge transformations of these two connections in the HGT
    formalism are given by the expressions:
        \bea
            \delta A&=&D_{A}\alpha-\tau({\lambda})\nonumber\\
            \delta\Sigma&=&D_{A}\lambda-\alpha\rhd\Sigma\label{eqn19}
        \eea
    where $\alpha$ is a $\textbf{g}$-valued 0-from gauge parameter and $\lambda$ is
    a $\textbf{h}$-valued 1-form gauge parameter.\\
\indent
    With these considerations we have finished the review of all the fundamental
    concepts of the HGT formalism that will the necessary for the generalization
    of (2+1)-dimensional gravity to the HGT formalism. However, before proceeding
    with the latter task, it is necessary to present a similar review of the
    $\Sigma\Phi EA$ model, since, as mentioned in the introduction, this model is
    the principal candidate for the generalization of (2+1)-dimensional gravity to
    the HGT framework.

\section{The $\Sigma\Phi EA$ model}

\indent
    In this section, we briefly review the basic aspects of the $\Sigma\Phi EA$ model
    as they are relevant to the generalization of the model to the HGT formalism. For
    more details about this model the interested reader is referred to \cite{mannpop}.
    In all of the following we will use, whenever possible, the same notation that we
    used in the previous section, in order to emphasize the similarities that exist
    between this model and the HGT formalism.\\
\indent
    As mentioned in the introduction, the action of the $\Sigma\Phi EA$ model is
    given by the expression:
        \be
            S_{\Sigma\Phi EA}=\int_{M}Tr\big\{E\wedge R[A]+
                        \Phi\wedge D_{A}\Sigma\big\}=
                        \int_{M}\big\{E_{i}\wedge R^{i}[A]+
                        \Phi_{i}\wedge D_{A}\Sigma^{i}\big\}\label{eqn20}
        \ee
    where $E$, $A$ are the $so(2,1)$-valued triad and (spin) connection 1-form fields,
    and $\Sigma$, $\Phi$ are $so(2,1)$-valued 2-form and respectively 0-form fields.
    The trace in (\ref{eqn20}) is the non-degenerate invariant bilinear form defined
    on the $so(2,1)$ Lie algebra through the relation:
        \be
            Tr(J^{i}J^{j})=\eta^{ij}\label{eqn21}
        \ee
    where $\{J^{i}, i=0,1,2\}$ are the generators of the $so(2,1)$ Lie algebra,
    satisfying the commutation relations:
        \be
            [J^{i},J^{j}]=\epsilon^{ijk}J_{k}\label{eqn22}
        \ee
    lower case latin indices from the middle of the alphabet are $so(2,1)$ algebra
    indices, $\eta=diag(-,+,+)$, and for the structure constants in (\ref{eqn22}) we
    use the convention $\epsilon^{012}=1$. The fields $\Sigma$ and $\Phi$ are coupled
    to gravity through the connection $A$ in the covariant derivative which is given
    by the expression:
        \be
            D_{A}\Sigma^{i}=d\Sigma^{i}+
                \epsilon^{ijk}A_{j}\wedge \Sigma_{k}\label{eqn23}
        \ee
\indent
    Up to surface terms, the first order variation of the action (\ref{eqn20}) yields
    the equations of motion:
         \bea
            R^{i}[A]&=&0\nonumber\\
            D_{A}E^{i}+\epsilon^{ijk}\Sigma_{j}\wedge \Phi_{k}&=&0\nonumber\\
            D_{A}\Sigma^{i}&=&0\label{eqn24}\\
            D_{A}\Phi^{i}&=&0\nonumber
        \eea
   and these equations of motion are invariant under the following infinitesimal gauge
   transformations:
        \bea
            \delta A^{i}&=&D_{A}\alpha^{i}\nonumber\\
            \delta \Phi^{i}&=&\epsilon^{ijk}\Phi_{j}\alpha_{k}\nonumber\\
            \delta \Sigma^{i}&=&D_{A}\lambda^{i}
                            -\epsilon^{ijk}\alpha_{j}\Sigma_{k}\label{eqn25}\\
            \delta E^{i}&=&D_{A}\beta^{i}+\epsilon^{ijk}(E_{j}\alpha_{k}
                            -\Phi_{j}\lambda_{k})\nonumber
        \eea
    with $\alpha^{i}$, $\beta^{i}$ 0-form and $\lambda^{i}$ 1-form $so(2,1)$-valued
    gauge parameters. Furthermore, the infinitesimal gauge transformations for the
    2-form field $\Sigma$ in (\ref{eqn25}) are themselves invariant under the
    infinitesimal "translations":
        \be
            \delta\lambda^{i}=D_{A}\rho^{i}\label{eqn26}
        \ee
    where now $\rho^{i}$ are 0-form $so(2,1)$-valued parameters, and under these
    circumstances, the number of independent gauge parameters describing the
    symmetries of the model reduces from 15 to 12.\\
\indent
    We conclude the review of the $\Sigma\Phi EA$ model with a few brief remarks
    regarding its solvability and solutions. Coupling the fields $\Sigma$, $\Phi$
    to pure gravity through the connection - as opposed to coupling them through
    the triads - has the main advantage that the resulting $\Sigma\Phi EA$ theory
    preserves the topological character inherited from pure gravity \cite{witten1},
    \cite{horowitz1}, \cite{frmapop}. Indeed, it can be shown \cite{mannpop} that
    the $\Sigma\Phi EA$ is a topological model with no local degrees of freedom,
    and as such it is solvable both classically and in the quantum framework.
    Consequently, its solutions are non-trivial for non-trivial topologies of the
    spacetime manifold $M$ and in the case of a spacetime manifold having the
    topology $M=\mathbb{R}\times S$, with $S$ a spacelike surface whose topology
    is that of a punctured plane, the model has as solutions the BTZ black-hole
    geometry \cite{mannpop} and the point particle geometry.\\
\indent
    We now have all the required mathematical and physical background for the
    generalization of (2+1)-dimensional pure gravity with vanishing cosmological
    to the HGT formalism. Furthermore, by now it should be quite clear not only
    that there are close similarities between the $\Sigma\Phi EA$ model and the
    HGT formalism, but also what these similarities are and how they could
    be exploited to reformulate the model in the HGT framework. In the next section,
    we will investigate in detail these similarities, and we will determine in detail
    a set of exact circumstances under which the $\Sigma\Phi EA$
    model becomes the sought HGT generalization of pure gravity in
    (2+1) dimensions.

\section{(2+1)-dimensional gravity as a higher gauge theory}

\indent
    As mentioned in Section I, with our choice of approach to this issue, generalizing
    (2+1)-dimensional gravity to the HGT formalism is equivalent with reformulating the
    $\Sigma\Phi EA$ model within the framework of HGT. In turn, this means on one hand
    that we must explicitly determine - in the differential picture - the components of
    the quadruple $(\textbf{g}, \textbf{h}, \tau, \rhd)$ involved in the definition of
    a Lie 2-algebra and on the other hand, that we must also determine a suitable action
    defined over this Lie 2-algebra which reduces to the action of the $\Sigma\Phi EA$
    model.\\
\indent
    Determining the Lie algebra $\textbf{g}$ is straightforward. Based on the fact that
    in pure (2+1)-dimensional gravity the 1-connection is a lorentzian connection, the
    obvious choice for $\textbf{g}$ is the Lie algebra $so(2,1)$. With this choice,
    and using the notation in Section III, the 1-connection of the HGT formalism becomes
    now an $so(2,1)$-valued connection 1-form $A=A_{i}J^{i}$.\\
\indent
    We still have to determine the remaining three components of the Lie 2-algebra
    quadruple - namely the Lie algebra $\textbf{h}$, the homomorphism $\tau$
    and the action$"\rhd"$ - and in order to do so, it is useful to summarize the
    relevant similarities between the $\Sigma\Phi EA$ model and the HGT formalism.
    The main reason behind this idea is that, as mentioned earlier, we can use
    these similarities as a guide to establish the remaining details of the
    reformulation of the $\Sigma\Phi EA$ model as an HGT.\\
\indent
    With the above choice for the Lie algebra $\textbf{g}$, the
    similarities between the two formulations can be summarized as
    follows:
        \begin{itemize}
            \item[\textbf{S1.}]
                Both the $\Sigma\Phi EA$ model and the HGT formalism involve a
                connection 1-form, defined on the Lie algebra
                $\textbf{g}\equiv so(2,1)$. Furthermore, they both impose strict
                conditions upon its SGT curvature, as given by (\ref{eqn16}) and
                (\ref{eqn24}).
            \item[\textbf{S2.}]
                Both the $\Sigma\Phi EA$ model and the HGT formalism involve a
                2-form $\Sigma$, which in the HGT formalism is a 2-connection, and
                as such it is defined on the Lie algebra \textbf{h} yet to be
                determined. If we ignore for the moment the explicit form of the
                action $"\rhd"$, the third equation of motion for the $\Sigma\Phi EA$
                model in (\ref{eqn24}) resembles very much a 2-flatness condition on
                the 3-curvature (\ref{eqn18}) of the 2-connection.
            \item[\textbf{S3.}]
                Once again, ignoring for the moment the explicit form of action
                $"\rhd"$ and of the homomorphism $\tau$, as it can be seen from
                (\ref{eqn19}) and (\ref{eqn25}), the infinitesimal gauge
                transformations for the 1-connection $A$ and the 2-form $\Sigma$ in
                both formulations are given by very similar expressions.
        \end{itemize}

\indent
    Consider now the observation ($\textbf{S2}$). It is obvious that if we want the
    equation of motion for $\Sigma$ in (\ref{eqn24}) to become a 2-flatness condition
    in the HGT formalism, we must define the action $"\rhd"$ of the Lie algebra
    $\textbf{g}$ on $\textbf{h}$ at the level of their generators as:
        \be
            J^{i}\rhd \chi^{m}=\epsilon^{im}_{\phantom{im}n}\chi^{n}\label{eqn27}
        \ee
    where for the generators of the Lie algebra $\textbf{h}$ we have used the notation
    in Section II, and $\epsilon^{im}_{\phantom{im}n}$ in the rhs of (\ref{eqn27}) is the totally
    antisymmetric 3-dimensional Levi-Civita symbol as defined in Section III.
    Furthermore, since the rhs of (\ref{eqn27}) is reminiscent of an adjoint action, in
    the following we will actually require that the action $"\rhd"$ be the adjoint
    action of the Lie algebra $\textbf{g}\equiv so(2,1)$ on $\textbf{h}$:
        \be
            J^{i}\rhd \chi^{m}=ad[J^{i}](\chi^{m})=[J^{i},\chi^{m}]=
                            \epsilon^{im}_{\phantom{im}n}\chi^{n}\label{eqn28}
        \ee
    This definition of the action $"\rhd"$ drastically restrict our choices for the Lie
    algebra $\textbf{h}$. According to (\ref{eqn28}) the algebra $\textbf{h}$ must be a
    3-dimensional algebra such that the adjoint action of $so(2,1)$ on $\textbf{h}$ is
    characterized by the Levi-Civita symbol $\epsilon^{ijk}$. Under these circumstances,
    the choices are obvious: the Lie algebra $\textbf{h}$ is either the Lorentz algebra
    $so(2,1)$ or the algebra of Poincar\'{e} translations $t^{3}$. Furthermore,
    it should be also noted that with the choice (\ref{eqn28}) for the action $"\rhd"$
    the infinitesimal gauge transformations for the 2-connection/2-form $\Sigma$ in both
    (\ref{eqn19}) and (\ref{eqn25}) become identical.\\
\indent
    Consider now the observation ($\textbf{S1}$). The restrictions imposed by the two
    formulations on the SGT curvature of the Lorentzian 1-connection $A$ are given
    by the relations:
        \bea
            R[A]&=&0\;\;(\Sigma\Phi EA)\nonumber\\
            R[A]&=&-\tau(\Sigma)\;\;(HGT)\label{eqn29}
        \eea
    It is clear that in order for these restrictions to match in both formulations,
    one should either have the 2-connection $\Sigma$ belonging to the kernel of the
    homomorphism $\tau$ or alternatively, one should have the homomorphism $\tau$ be
    the trivial homomorphism $\tau(Y)=0$ for any $Y\in \textbf{h}$.\\
\indent
    Similarly, if one considers the observation ($\textbf{S3}$), then the infinitesimal
    gauge transformations for the Lorentzian 1-connection $A$ are given by the
    expressions:
        \bea
            \delta A&=&D_{A}\alpha \;\;(\Sigma\Phi EA)\nonumber\\
            \delta A&=&D_{A}\alpha-\tau(\lambda)\;\;(HGT)\label{eqn30}
        \eea
    and once again, in order to have the two relations match, one should either have
    the gauge parameter $\lambda$ belonging to the kernel of the homomorphism $\tau$
    or alternatively, one should have the homomorphism $\tau$ be the trivial
    homomorphism.\\
\indent
    With these considerations, we can now return to the issue of determining the Lie
    algebra $\textbf{h}$ of the Lie 2-algebra quadruple. It is straightforward to see
    that the Lie algebra $\textbf{h}$ cannot be $so(2,1)$, since if that were the case,
    the quadruple $\big(so(2,1), so(2,1), \tau, ad[so(2,1)](so(2,1))\big)$ would not
    satisfy the requirements of a Lie 2-algebra. Explicitly, it is the second condition
    in (\ref{eqn11}), namely:
        \be
          \tau(Y)\rhd Y'=[Y,Y']\label{eqn31}
        \ee
    for any $Y, Y'\in so(2,1)$ that is not satisfied by the 2-connection by the
    2-connection $\Sigma$ and its 1-form gauge parameter $\lambda$. The reason
    why these two forms do not satisfy (\ref{eqn31}) are obvious: according to
    this condition, if $\Sigma$, $\lambda$ belong to the kernel of the homomorphism
    $\tau$, they must also belong to the center of the Lie algebra
    $\textbf{h}\equiv so(2,1)$, and $so(2,1)$ has only a trivial center. This would
    imply in turn that the 2-connection and its gauge parameter can only be null forms,
    which is not a useful result for our purposes.\\
\indent
    On the other hand, the condition (\ref{eqn31}) is identically satisfied for $\Sigma$
    and $\lambda$ in the kernel of $\tau$ if $\textbf{h}\equiv t^{3}$,  since
    the algebra of Poincar\'{e} translations is abelian. Furthermore, in this case
    (\ref{eqn31}) together with (\ref{eqn28}) implies that in fact the homomorphism
    $\tau$ can only be the trivial homomorphism $\tau=0$.\\
\indent
    Under these circumstances, the appropriate HGT mathematical structure underlying
    the generalization of (2+1)-dimensional gravity to the HGT formalism is in the
    differential picture the Lie 2-algebra given by the quadruple
    $\big(so(2,1), t^{3}, \tau=0,\, ad[so(2,1)](t^{3})\big)$, i.e. the adjoint or
    tangent Poincar\'{e} 2-algebra \cite{Baez1}, \cite{PfGir}.\\
\indent
    It is useful at this time to give a brief summary of the above considerations.
    According to the previous arguments, the $\Sigma\Phi EA$ model could be reformulated
    as an HGT with the Poincar\'{e} 2-algebra. In this case, the two connections would
    satisfy the equations of motion:
        \bea
            R[A]&=&0\nonumber\\
            G[\Sigma]&=&0\label{eqn32}
        \eea
    i.e. the reformulation implies an HGT that is simultaneously 1-flat and 2-flat, and
    they should be invariant under the infinitesimal gauge transformations:
        \bea
            \delta A&=&D_{A}\alpha\nonumber\\
            \delta\Sigma&=&D_{A}\lambda-\alpha\rhd\Sigma\label{eqn33}
        \eea
    with $\alpha$ an $so(2,1)$-valued 0-form gauge parameter and $\lambda$ a
    $t^{3}$-valued 1-form parameter. Furthermore, due to the 1-flatness of the
    model, the gauge transformation for the 2-connection $\Sigma$ is itself invariant
    under the infinitesimal "translation":
        \be
            \delta\lambda=D_{A}\rho\label{eqn34}
        \ee
    where $\rho$ is a $t^{3}$-valued 0-form parameter.\\
\indent
    What still remains to be done is to find an HGT action which upon first order
    variation yields as equations of motion the equations (\ref{eqn32}) for the
    2-connections and the remaining equations in (\ref{eqn24}), and which is
    invariant under the infinitesimal transformations (\ref{eqn33}), (\ref{eqn34})
    and under the remaining gauge transformations in (\ref{eqn25}). Explicitly,
    and based on the considerations in Section I regarding the generalized form of
    the HGT action, this means that we need to assign the remaining fields $E$ and
    $\Phi$ to the appropriate Lie algebras of the 2-algebra quadruple, and that we
    must also determine an appropriate invariant trace operator for the action
    integral.\\
\indent
    The easiest way to solve these remaining problems is to first consider the issue
    of the trace operator in the action integral. As mentioned in the introduction,
    the generalized action should be given by an expression of the form:
        \be
            S_{HGT}=\int_{M}\big\{Tr_{1}\big(E\wedge R[A]\big)+
                    Tr_{2}\big(\Phi\wedge G[\Sigma]\big)\big\}\label{eqn35}
        \ee
    and we already know that the HGT theory should be formulated with the Poincar\'{e}
    2-algebra such that the 1-connection is an $so(2,1)$-valued 1-form and the
    2-connection is a $t^{3}$-valued 2-form. Furthermore, it is known
    \cite{witten1} that on the Poincar\'{e} Lie algebra one can define two
    non-degenerate invariant bilinear forms, given by the expressions:
        \bea
            Tr_{\textbf{g}}(J^{i}J^{j})=Tr_{\textbf{h}}(P^{i}P^{j})&=&\eta^{ij}\nonumber\\
            Tr(J^{i}P^{j})&=&0\label{eqn36}
        \eea
    and
        \bea
            Tr_{\textbf{P}}(J^{i}J^{j})=Tr_{\textbf{P}}(P^{i}P^{j})&=&0\nonumber\\
            Tr_{\textbf{P}}(J^{i}P^{j})&=&\eta^{ij}\label{eqn37}
        \eea
    where in (\ref{eqn36}) and (\ref{eqn37}) we have used the traditional notation
    for the generators of the Poincar\'{e} algebra, and we have used the formal
    notation $\textbf{g}\equiv so(2,1)$ and $\textbf{h}\equiv t^{3}$. Under these
    circumstances, we can only have two choices for assigning the remaining fields
    $E$ and $\Phi$ the algebras of the 2-algebra quadruple:
        \begin{itemize}
            \item[\textbf{A.}]
                If we use the bilinear form (\ref{eqn36}), then $E$ must be an
                $\textbf{g}$-valued 1-form, $\Phi$ must be a  $\textbf{h}$-valued
                0-form, and the generalized action will be given by the formal
                expression:
                    \be
                        S_{\textbf{HGT-A}}=\int_{M}\big\{Tr_{\textbf{g}}\big(E\wedge R[A]\big)+
                        Tr_{\textbf{h}}\big(\Phi\wedge G[\Sigma]\big)\big\}\label{eqn38}
                    \ee
            \item[\textbf{B.}]
                If we use the bilinear form (\ref{eqn37}), then $E$ must be
                an
                $\textbf{h}$-valued 1-form, $\Phi$ must be a $\textbf{g}$-valued 0-form,
                and the generalized action will be given by the formal expression:
                    \be
                        S_{\textbf{HGT-B}}=\int_{M}Tr_{\textbf{P}}\big\{E\wedge R[A]+
                                    \Phi\wedge G[\Sigma]\big\}\label{eqn39}
                    \ee
        \end{itemize}

\indent
    Of course, both these actions yield the same equations of motion and have the same
    gauge symmetries as the $\Sigma\Phi EA$ model. As such they represent the
    HGT reformulation(s) of this model, and hence the sought generalization(s) of
    (2+1)-dimensional pure gravity with vanishing cosmological constant to the HGT
    formalism. The only difference between the two generalization is that for a
    vanishing 2-connection, the actions (\ref{eqn38}) and (\ref{eqn39}) reduce to
    pure gravity as a gauge theory formulated over the $so(2,1)$ Lie algebra and
    respectively over the Poincar\'{e} algebra.\\
\indent
    We conclude this section with a few remarks regarding the non-triviality of these
    generalizations. As mentioned earlier, as a topological model the $\Sigma\Phi EA$
    model has non-trivial solutions for non-trivial topologies, and as such one would
    expect the above HGT generalization(s) to behave similarly since they are just
    reformulation(s) of the model in different framework. In other words, we expect
    that the HGT generalization(s) of our model - and in general any HGT with no local
    degrees of freedom - should also have non-trivial solutions for non-trivial
    topologies of the spacetime manifold $M$, and while ultimately this is indeed the
    case, it is useful at this time to analyze in more detail the circumstances under
    which this statement is valid.\\
\indent
    To better illustrate the issue that we intend to address, it is useful to revert back
    to the integral picture of the HGT formalism and recall the labeling procedure for
    curves and surfaces in the spacetime manifold. In order to preserve the analogy
    with the SGT formulation of the $\Sigma\Phi EA$ model, we will confine our
    considerations to the spatial leafs $S$ of the foliation of the spacetime $M=R\times S$.
    Under these circumstances, the labeling procedure for curves and surfaces embedded in
    $S$ can be summarized as follows: if $P,Q\in S$ are two distinct points in S and we
    connect these points by two distinct curves $\gamma_{1}$ and $\gamma_{2}$ such that
    the surface bound by these curves has the topology of a disc \cite{PfGir}, then we
    can label the two curves by the holonomies $g_{1},g_{2}\in G$ of the 1-connection
    and the surface bound by these curves by a group element $h\in H$ which can be
    interpreted as the analog of the Wilson line for the 2-connection. In the HGT
    framework, these labels are constrained to obey the condition (\ref{eqn8}), which
    can be rewritten in the equivalent form \cite{PfGir}:
        \be
            t(h)=g_{2}\cdot g_{1}^{-1}\label{eqn40}
        \ee
    It is useful to note that in this form, and as far as the labeling process is concerned,
    (\ref{eqn40}) can be given the following interpretation: surfaces are labeled by group
    elements $h\in H$ associated through the homomorphism $t$ to the (inverse) holonomy of
    the 1-connection along the boundary of the surface. Loosely speaking, the homomorphism
    $t$ associates to each 1-holonomy of the 1-connection a 2-holonomy of the 2-connection.\\
\indent
    As it clear from the above considerations, the labeling procedure relies heavily
    on the requirement that the surface bound by the source and target curves $\gamma_{1}$
    and respectively $\gamma_{2}$ should have the topology of a disk. In a sense, this
    requirement is not at all surprising, since in fact the group elements $h\in H$ labeling
    surfaces are in fact given by (the exponential of) a surface integral of the 2-connection
    \cite{PfGir}. However, while this requirement is definitely compatible with a trivial
    topology of the spatial surface $S$, it is not at all clear if and how the requirement -
    and hence the condition (\ref{eqn40}) - is also compatible with a non-trivial topology
    of $S$. This is the issue that needs to be addressed in order to establish whether
    or not the HGT generalizations of the $\Sigma\Phi EA$ model admit non-trivial solutions.
    In particular, and since our immediate purpose is to answer the question of whether or
    not these generalizations admit the same solutions as the SGT formulation of the model,
    we will confine our considerations only to the case where the surface $S$ has the
    topology of a punctured plane. However, for clarity purposes, it is more convenient to
    develop our arguments for a surface $S$ having the topology of an open annulus. Since
    the open annulus and the punctured plane are homeomorphic and hence homotopically
    equivalent, the conclusions of the analysis developed for the former (which is easier
    to visualize) apply identically to the latter.\\
\indent
    Under these circumstances, if we consider now the annular topology, the fundamental
    group of this surface contains two principal classes of loops, namely the class of
    contractible loops (homotopic to the null loop) and the class of non-contractible
    loops surrounding the inner circumference of the annulus\footnote{For the purpose
    of the present argument, it suffices to consider only the class of loops surrounding
    one time the inner circumference of the annulus}. For the class of contractible loops,
    the surfaces bound by these loops have the topology of a disc, and as such the
    condition (\ref{eqn40}) is satisfied, albeit in a trivial manner.\\
\indent
    For the class of non-contractible loops surrounding the inner circumference of the
    annulus, the situation is quite different. Such loops do not enclose a surface,
    since such a surface simply does not exist (it is exactly the surface that has been
    "removed" from the open disk in order to create the annular surface), and hence there
    exists no label $h\in H$ that can be associated to the holonomies of the 1-connection
    along these loops by the homomorphism $t$ in accordance to the prescription described
    earlier. The obvious conclusion to these arguments is that for such non-contractible
    loops the condition (\ref{eqn40}) becomes invalid. Loosely speaking once again, this
    conclusion states that there exists no 2-holonomy $h\in H$ that can be associated to
    the 1-holonomies corresponding to this class of loops through the homomorphism $t$.
    And as mentioned above, this conclusion is also valid for the punctured plane
    topology.\\
\indent
    Far from being an inconvenient result, the above conclusion solves in fact the
    issue of whether or not the HGT generalizations of the $\Sigma\Phi EA$ model
    admit non-trivial solutions for a surface $S$ having the topology of a punctured
    plane or an annulus. We can state at this time that the HGT generalizations of
    the $\Sigma\Phi EA$ model - and HGT in general - are compatible with a non-trivial
    topology of the open surface $S$ (i.e. have non-trivial solutions) provided that
    (\ref{eqn40}) is valid for all the contractible curves in $S$ and provided that
    there exist non-contractible loops in $S$ for which the condition (\ref{eqn40})
    fails to be valid in the sense discussed above\footnote{It must be emphasized
    that this statement regarding the existence of non-trivial solutions of HGT
    for non-trivial topologies of the surface $S$ is only valid if $S$ is an open
    surface. If $S$ is a closed surface, the statement becomes invalid since for
    example if $S$ has the topology of a torus, it is straightforward to see that
    the condition (\ref{eqn40}) remains valid for both non-contractible curves
    surrounding one time the large and small circumferences of the
    torus.}.\\
\indent
    We can conclude based on the above considerations that, as stated earlier, the
    HGT generalizations of the $\Sigma\Phi EA$ model developed in this section are
    non-trivial in the sense that they are compatible with the punctured plane
    topology of the surface $S$. Furthermore, it becomes now clear that for this
    topology of $S$ they also admit the same solutions (e.g. the point-particle
    solution and the BTZ black-hole solution) as the SGT formulation of the model
    \cite{mannpop}.

\section{Conclusions}

\indent
    In this paper we have considered the issue of the generalization of
    (2+1)-dimensional gravity to the HGT formalism, and we have shown that one such
    possible generalization arises from the reformulation of the $\Sigma\Phi EA$
    model as a higher gauge theory. The generalized theory is an HGT with the
    adjoint Poincar\'{e} 2-group, and it inherits from the $\Sigma\Phi EA$ model
    its non-trivial solutions that include the BTZ black-hole geometry and the
    point-particle geometry.\\
\indent
    While it is very interesting that the HGT formalism can also find its
    application to the study of classical gravity - in (2+1) dimensions at
    least - this is just a first simple step in exploring its potential and
    implications in this latter context. Further investigation is required in
    order to determine if more complex such generalizations of classical gravity
    can be developed in (2+1) dimensions (e.g. generalizations involving non-abelian
    2-connections, non-vanishing cosmological constants) and also if the formalism
    can be successfully applied to the more realistic theory of gravity in (3+1)
    dimensions.\\
\indent
    Nevertheless, even in the context of (2+1)-dimensional gravity, the existence of
    an HGT generalization of pure gravity with vanishing cosmological constant opens
    a whole new range of issues that are worth exploring. In particular, it would be
    extremely interesting to explore the differences between the spin-foam quantum
    theories of the $\Sigma\Phi EA$ model as a traditional SGT and as an
    HGT, and we hope to be able to further study this issue in a
    future paper.\\

\noindent
    \textbf{Acknowledgements}\\
    \\
\indent
    This work was supported in part by the Natural Sciences and Engineering Research
    Council of Canada. The authors would also like to thank Florian Girelli for many
    useful discussions on 2-groups and 2-gauge theory.

\end{document}